\documentstyle[12pt]{article}

\begin{document}
\baselineskip=20.5pt

\def\beqra{\begin{eqnarray}} \def\eeqra{\end{eqnarray}}
\def\beqast{\begin{eqnarray*}} \def\eeqast{\end{eqnarray*}}
\def\beq{\begin{equation}}	\def\eeq{\end{equation}}

\def\fnote#1#2{\begingroup\def\thefootnote{#1}\footnote{#2}\addtocounter
{footnote}{-1}\endgroup}

\def\ut#1#2{\hfill{UTTG-{#1}-{#2}}}
\def\uw#1#2{\hfill{WISC-MILW-{#2}-TH-{#1}}}

\def\sppt{Research supported in part by the
Robert A. Welch Foundation and NSF Grant PHY 9009850}

\def\utgp{\it Theory Group\\ Department of Physics \\ University of Texas
\\ Austin, Texas 78712}

\def\gam{\gamma}
\def\Gam{\Gamma}
\def\la{\lambda}
\def\eps{\epsilon}
\def\La{\Lambda}
\def\si{\sigma}
\def\Si{\Sigma}
\def\al{\alpha}
\def\bet{\beta}
\def\Tha{\Theta}
\def\tha{\theta}
\def\vphi{\varphi}
\def\del{\delta}
\def\Del{\Delta}
\def\ab{\alpha\beta}
\def\om{\omega}
\def\Om{\Omega}
\def\mn{\mu\nu}
\def\mun{^{\mu}{}_{\nu}}
\def\kap{\kappa}
\def\rsi{\rho\sigma}
\def\beal{\beta\alpha}

\def\til{\tilde}
\def\rta{\rightarrow}
\def\eqv{\equiv}
\def\nab{\nabla}
\def\pa{\partial}
\def\sit{\tilde\sigma}
\def\ul{\underline}
\def\indt{\parindent2.5em}
\def\nd{\noindent}

\def\rsi{\rho\sigma}
\def\beal{\beta\alpha}
\def\pha{\phi^{\al}_E}
\def\phb{\phi^{\bet}_E}

\def\caa{{\cal A}}
\def\cb{{\cal B}}
\def\cac{{\cal C}}
\def\cd{{\cal D}}
\def\ce{{\cal E}}
\def\cf{{\cal F}}
\def\cg{{\cal G}}
\def\cah{{\cal H}}
\def\ci{{\cal I}}
\def\cj{{\cal{J}}}
\def\ck{{\cal K}}
\def\cl{{\cal L}}
\def\cm{{\cal M}}
\def\cn{{\cal N}}
\def\cO{{\cal O}}
\def\cp{{\cal P}}
\def\car{{\cal R}}
\def\cs{{\cal S}}
\def\ct{{\cal{T}}}
\def\cu{{\cal{U}}}
\def\cv{{\cal{V}}}
\def\cw{{\cal{W}}}
\def\cx{{\cal{X}}}
\def\cy{{\cal{Y}}}
\def\cz{{\cal{Z}}}

\def\raisenot{\raise .5mm\hbox{/}}
\def\nota{\ \hbox{{$a$}\kern-.49em\hbox{/}}}
\def\notA{\hbox{{$A$}\kern-.54em\hbox{\raisenot}}}
\def\notb{\ \hbox{{$b$}\kern-.47em\hbox{/}}}
\def\notB{\ \hbox{{$B$}\kern-.60em\hbox{\raisenot}}}
\def\notc{\ \hbox{{$c$}\kern-.45em\hbox{/}}}
\def\notd{\ \hbox{{$d$}\kern-.53em\hbox{/}}}
\def\notbd{\ \hbox{{$D$}\kern-.61em\hbox{\raisenot}}} 
\def\note{\ \hbox{{$e$}\kern-.47em\hbox{/}}}
\def\notk{\ \hbox{{$k$}\kern-.51em\hbox{/}}}
\def\notp{\ \hbox{{$p$}\kern-.43em\hbox{/}}}
\def\notq{\ \hbox{{$q$}\kern-.47em\hbox{/}}}
\def\notW{\ \hbox{{$W$}\kern-.75em\hbox{\raisenot}}}
\def\notz{\ \hbox{{$Z$}\kern-.61em\hbox{\raisenot}}}
\def\notpa{\hbox{{$\partial$}\kern-.54em\hbox{\raisenot}}}

\def\fo{\hbox{{1}\kern-.25em\hbox{l}}}  
\def\rf#1{$^{#1}$}
\def\bx{\Box}
\def\tr{{\rm Tr}}
\def\rmtr{{\rm tr}}
\def\dgg{\dagger}

\def\lag{\langle}
\def\rag{\rangle}
\def\bmid{\big|}

\def\vlap{\overrightarrow{\La p}} 
\def\lrta{\longrightarrow} \def\lrar{\raisebox{.8ex}{$\longrightarrow$}}
\def\rlarw{\longleftarrow\!\!\!\!\!\!\!\!\!\!\!\lrar}

\def\llra{\relbar\joinrel\longrightarrow}              
\def\mapright#1{\smash{\mathop{\llra}\limits_{#1}}}    
\def\mapup#1{\smash{\mathop{\llra}\limits^{#1}}}     

\def\7#1#2{\mathop{\null#2}\limits^{#1}}	
\def\5#1#2{\mathop{\null#2}\limits_{#1}}	
\def\too#1{\stackrel{#1}{\to}}
\def\tooo#1{\stackrel{#1}{\longleftarrow}}
\def\nout{{\rm in \atop out}}

\def\one{\raisebox{.5ex}{1}}
\def\BM#1{\mbox{\boldmath{$#1$}}}

\def\ltsim{\matrix{<\cr\noalign{\vskip-7pt}\sim\cr}}
\def\gtsim{\matrix{>\cr\noalign{\vskip-7pt}\sim\cr}}
\def\haf{\frac{1}{2}}


\def\place#1#2#3{\vbox to0pt{\kern-\parskip\kern-7pt
                             \kern-#2truein\hbox{\kern#1truein #3}
                             \vss}\nointerlineskip}

\def\illustration #1 by #2 (#3){\vbox to #2{\hrule width #1 height 0pt depth
0pt
                                       \vfill\special{illustration #3}}}

\def\scaledillustration #1 by #2 (#3 scaled #4){{\dimen0=#1 \dimen1=#2
[B           \divide\dimen0 by 1000 \multiply\dimen0 by #4
            \divide\dimen1 by 1000 \multiply\dimen1 by #4
            \illustration \dimen0 by \dimen1 (#3 scaled #4)}}

\def\ON{{\cal O}(N)}
\def\UN{{\cal U}(N)}
\def\bdPh{\mbox{\boldmath{$\dot{\!\Phi}$}}}
\def\bPh{\mbox{\boldmath{$\Phi$}}}
\def\bPhs{\bPh^2}
\def\sef{S_{eff}[\sigma]}
\def\sigx{\sigma(x)}
\def\bph{\mbox{\boldmath{$\phi$}}}
\def\bphs{\bph^2}
\def\ex{\BM{x}}
\def\exs{\ex^2}
\def\xdot{\dot{\!\ex}}
\def\y{\BM{y}}
\def\ys{\y^2}
\def\ydot{\dot{\!\y}}
\def\pat{\pa_t}
\def\pax{\pa_x}

\renewcommand{\thesection}{\Roman{section}}

\ut{5}{95}

\uw{12}{95}

\hfill{hep-th/yymmdd}

\vspace*{.1in}
\begin{center}
  \large{\bf Self-Adjoint Wheeler-DeWitt Operators, the Problem of Time
and the Wave Function of the Universe}
\normalsize

\vspace{18pt}
Joshua Feinberg\fnote{*}{ Supported by the Robert A. Welch Foundation and NSF
Grant PHY 9009850.}

\vspace{7pt}
{\it Theory Group, Department of Physics\\
The University of Texas at Austin, RLM5.208, Austin, Texas 78712\\
\vspace{4pt}
e-mail joshua@utaphy.ph.utexas.edu}

\vspace{10pt}
Yoav Peleg\fnote{**}{Supported by the NSF grants PHY-9105935 and
PHY-9315811.}

\vspace{7pt}
{\it Department of Physics\\
University of Wisconsin at Milwaukee, Milwaukee, WI 53201\\
\vspace{4pt}
e-mail yoav@alpha2.csd.uwm.edu}

\vspace{.5cm}

\end{center}

\begin{minipage}{5.3in}
{\abstract~~~~~
We discuss minisuperspace aspects a non empty Robertson-Walker universe
containing scalar matter field. The requirement that the Wheeler-DeWitt (WDW)
operator be self adjoint is a key ingredient in constructing the physical
Hilbert space and has non-trivial cosmological implications since it is related
with the problem of time in quantum cosmology. Namely, if time is parametrized
by matter fields we find two types of domains for the self adjoint WDW
operator:
a non trivial domain is comprised of zero current (Hartle-Hawking type) wave
functions and is parametrized by two new parameters, whereas the domain of a
self adjoint WDW operator acting on tunneling (Vilenkin type) wave functions is
a {\em single} ray. On the other hand, if time is parametrized by the scale
factor both types of wave functions give rise to non trivial domains for the
self adjoint WDW operators, and no new parameters appear in them.}

\end{minipage}

\vspace{15pt}

\vfill
\pagebreak

\newcommand{\be}{\begin{equation}}
\newcommand{\ee}{\end{equation}}
\renewcommand{\thesection}{\arabic{section}}
\renewcommand{\theequation}{\thesection.\arabic{equation}}

\section{Introduction}
\setcounter{equation}{0}

One of the simplest models of quantum cosmology is the
Robertson-Walker (RW) minisuperspace. RW geometries describe homogeneous and
isotropic universes. The RW geometry is defined by the line element
  \be
ds^{2} = - N_{\perp}^{2} d\eta^{2} + a^{2}(\eta) d\Omega_{3}^{2}  .
\label{RW}
 \ee

In (\ref{RW}) the only dynamical degree of freedom is the scale factor
$a(\eta)$. The lapse function $N_{\perp}$ is not dynamical , being a pure gauge
variable. The quantity $d\Omega_{3}^2$ is the standard line element on the
unit three-sphere. We use units in which $\hbar = c = 1$ and $G = M_p^{-2}
= 3\pi / 4$.

The pure gravitational action corresponding to (\ref{RW}) is
  \begin{eqnarray}
S_{g} &=& \frac{1}{16\pi G} \int_{M} d^{4}x \sqrt{-g} \left( R
- 2 \Lambda \right) + \frac{1}{8\pi G} \int_{\partial M} d^{3}x
\sqrt{g^{(3)}} K \nonumber \\
&=& \int d\eta N_{\perp} a^{3} \left[
a^{-2} \left( 1 - \frac{\dot{a}^{2}}{N_{\perp}^{2}} \right)
- \frac{\Lambda}{3} \right]  .
\label{1.2}
  \end{eqnarray}
In (\ref{1.2}) $\Lambda$ is the cosmological constant\footnote{One can
regard $\Lambda$ as a pure cosmological constant or as the vacuum energy
of some non trivial field configurations, or a sum of both.},
$M=I \times S^{3}$ is the space-time manifold, $K=K^i_i$ is the trace of the
second fundamental form of the space-like boundary $\partial M = S^{3}$
(defined by $\eta=const.$)
and the dot denotes differentiation with respect to $\eta$.
The Hamiltonian corresponding to (\ref{1.2}) is
  \be
H = - N_{\perp} \left(  \frac{1}{4a} P_{a}^{2} + a - g^{2} a^{3}
\right)
\label{1.3}
  \ee
where $P_{a} = \partial L / \partial \dot{a} = - 2a \dot{a}/N_{\perp}$
is the canonical momentum conjugate to $a(\eta)$ and
$g^{2} = \frac{\Lambda}{3}$. It is assumed in (\ref{1.3}) that $\Lambda
\geq 0$. Gauge invariance of (\ref{1.2}) yields the Hamiltonian constraint
\be
- {\partial H \over \partial N_{\perp}} = \frac{1}{4a} P_{a}^{2} + a - g^{2}
a^{3} = 0 .
\label{1.4}
\ee
The constraint (\ref{1.4}) requires a gauge fixing condition. A possible such
gauge fixing condition is $ N_{\perp}=const.\neq 0$ in which the time variable
$\eta$ becomes essentially the proper time $\tau$. In this gauge the solution
of the classical equations of motion (with initial conditions $a(0)=g^{-1}$,
$\dot{a}(0)=0$) is
  \be
a(\tau) = g^{-1} \mbox{cosh}(g\tau) ~,
\label{1.5}
\ee
which describes a universe that contracts from an infinite
radius in the absolute past, reaches a minimum radius, $a_{min}
= g^{-1}$, and re-expands to infinity in the absolute future.

Quantization of this simple system is accomplished straightforwardly in the
coordinate representation by the usual operator realizations
  \be
\hat{a} = a ~~~~~\mbox{and}~~~~~ \hat{P}_{a} = -i\frac{
\partial}{\partial a}  .
\label{1.6}
  \ee
\par
Neglecting operator ordering problems in the kinetic term\footnote{In this
work we study only semiclassical solutions to the WDW equation,
where operator ordering issues are not so important.} the Hamiltonian
constraint becomes the Wheeler-DeWitt (WDW) equation
\cite{Wheeler,DeWitt} for the wave function of the Universe:
  \be
\left( - \frac{1}{4a} \frac{\partial^{2}}{\partial
a^{2}} + a - g^{2} a^{3} \right) \Psi(a) = 0 ~.
\label{1.7}
  \ee
Equation (\ref{1.7}) is a Schr\"{o}dinger equation $\hat{H} \Psi = 0$ for a
zero energy eigenstate of a mass $m=2$ particle, moving in the one
dimensional potential
  \be
V(a) = a^{2} - g^{2} a^{4} .
\label{1.8}
  \ee

Despite the fact that the potential (\ref{1.8}) is unbounded from below, it has
the property that the time of flight of a classical particle from the largest
turning point to infinity {\em is finite}, namely,
  \be
\int_{x_{0} > {1\over g}}^{\infty} \frac{dx}{\sqrt{|V(x)|}} < \infty \,.
\label{1.9}
  \ee
This may seem contradictory to (\ref{1.5}) at first sight , because this
equation implies $\tau \rightarrow \infty$ as $a \rightarrow \infty$. However,
this contradiction is only apparent, since what we call ``time" in (\ref{1.9})
is not the proper time $\tau$. In order that the classical particle that moves
in the one dimensional potential (\ref{1.8}) have the standard kinetic energy
term (that is an implicit assumption in (\ref{1.9})), one must impose
the ``conformal time" gauge  $N_{\perp}=a(t)$ on (\ref{1.3}). Doing so,
(\ref{1.3}) becomes
 \begin{equation}
H = - \left( \frac{P_{a}^{2}}{4} + a^{2} - g^{2} a^{4}  \right).
\label{1.10}
\end{equation}
The functional relation between the conformal time and proper time is
  \be
t = \int \frac{d\tau}{a(\tau)} = \mbox{tan}^{-1}\left( \mbox{
sinh}(g\tau) \right)
\label{1.11}
  \ee
and the classical solution (\ref{1.5}) in the conformal gauge is
  \be
a(t) = \frac{1}{g \mbox{cos}(t)} ~.
\label{1.12}
  \ee
It is clear from (\ref{1.12}) that the ``particle"  reaches
infinity indeed after finite conformal time, $t=\pi/2$.

Eq. (\ref{1.9}) suggests that one dimensional Schr\"{o}dinger operators like
(\ref{1.10}) are very similar to
Schr\"{o}dinger operators describing quantal systems defined on a finite
segment of the real line, despite the fact that they act on wave functions
supported along the real positive half line. Quantal systems defined on a
finite segment require boundary conditions as an essential part of their
definition. In a similar manner, Schr\"{o}dinger operators like (\ref{1.10})
require boundary conditions on wave functions, ensuring probability
conservation at infinity. Thus, such conditions extend ill defined Hamiltonians
like (\ref{1.10}) into a self-adjoint form which generate unitary time
evolution operators \cite{von-Neumann,Reed-Simon,Farhi1,Farhi3,Farhi2}.
\par
Considerations of extending the Hamiltonian (\ref{1.10}) into a self-adjoint
form seem irrelevant for quantum cosmological considerations at first sight.
Indeed, solutions of the WDW equation (\ref{1.7}) for the pure RW
geometry (\ref{RW}) are always the {\em zero} energy eigenstate of (\ref{1.10})
which spans a one dimensional Hilbert space on which (\ref{1.10}) is trivially
self adjoint\footnote{Note that the
usual arguments that guarantee non-degeneracy of the discrete spectrum of one
dimensional Schr\"odinger operators, and in particular-that the corresponding
wave function are real up to an overall phase, are inapplicable for the
operator in (\ref{1.10}) which is unbounded from below. We elaborate on this
point in section 3.}. This is, however, only an incorrect superficial
statement.
There are at least two important reasons to introduce self-adjoint
extensions of the WDW Hamiltonian of which (\ref{1.10}) is only a
very simple case. First, we note that the null condition (\ref{1.4})
is just a special case of the general constraint enforcing reparametrization
invariance on {\sl physical states\/} in quantum gravity, namely, that they be
annihilated by the WDW operator. This implies that one has to
consider the WDW operator defined in a Hilbert space larger than
the space of physical states, where one can apply it on various states and
check whether they are physical or not. In this framework of constrained
quantization it is always assumed that {\it all} relevant operators, and in
particular, the constraint operators, are self-adjoint with respect to the
Hilbert space inner product\cite{Henneaux}. In our particular case, this means
that we have to include in the domain of definition of the operator
(\ref{1.10})
many eigenstates with non vanishing energy eigenvalues. In this case, the
issue of a proper self-adjoint extension of (\ref{1.10}) becomes important.

Second, note that when cases of a non-empty RW universes or perturbed RW
universes \cite{perturbed,Vilen2} are considered, many non zero
energy eigenstates of (\ref{1.10})
become relevant, even in the {\sl physical\/} Hilbert subspace itself.
This comes about because now the WDW constraint implies that the
{\em total} Hamiltonian must annihilate physical states. To see this we observe
that the total Hamiltonian may be written as
  \be
H_{WDW}\equiv H_{tot} =  H_{\phi,h_{\mu \nu}} - H_{0}
\label{1.13}
  \ee
where $ - H_{0}$ is (\ref{1.10}) and $H_{\phi,h_{\mu \nu}}$ is
the Hamiltonian of matter fields (denoted here by the field $\phi$)
and of gravitational perturbations (denoted by $h_{\mu \nu}$)\footnote{That is,
the metric is $g_{\mu \nu} = g^{(0)}_{\mu \nu}+ h_{\mu \nu}$ where $g^{(0)}$ is
the RW metric in (\ref{RW}).}. The minus sign in (\ref{1.13}) results from the
fact that the kinetic term of the conformal mode $a$ in (\ref{1.2}) appears
with
the ``wrong" sign. In simple cases where a separation of variables applies,
$|\psi\rangle = |\phi_{a}\rangle |\chi_{\phi,h} \rangle$, the corresponding
WDW equation reduces into the two equations
  \be
\hat{H}_{0} | \phi_{a} \rangle = E | \phi_{a} \rangle \\
\label{1.14}
  \ee
and
  \be
\hat{H}_{\phi,h} |\chi_{\phi,h}\rangle  = E |\chi_{\phi,h}\rangle
\label{1.15}
  \ee
where $E$ is some non-negative eigenvalue\footnote{The requirement that
$E\geq 0$ is a consequence of the energy condition for matter fields.} either
of $\hat{H}_{0}$ or of $\hat{H}_{\phi,h}$. Now, in principle, many eigenstates
of $\hat{H}_0$ and of
$\hat{H}_{\phi,h}$ (sharing the same eigenvalue) may enter into the quantum
state of the Universe, especially if it turns out to be a quasi-classical
state\footnote{ A quasi-classical state is a superposition of many energy
eigenstates which are superimposed with some well behaved amplitudes, for
example,
  \begin{eqnarray*}
\Psi_{qc}(a) \sim \int \Psi_{E}(a) \mbox{exp}\left[-\left(\frac{E-E_{0}}
{\Delta E} \right)^2\right] dE ~,
  \end{eqnarray*}
where $\Psi_{E}$ are solutions of (\ref{1.14}) and
$E_{0}$ is the classical energy.}.

Now that many different energy eigenstates of (\ref{1.14}) and (\ref{1.15})
are used to construct physical states, one has to equip them with a
{\em time independent} inner product in order to define the physical Hilbert
space. We find that if we choose matter fields as our clock,
self-adjointness of (\ref{1.10}) turns out to be the necessary and sufficient
condition for the existence of such an inner product. In this case there is a
continuous two parameter family of non trivial domains of the self-adjoint WDW
operator that are spanned by zero current (Hartle-Hawking type) wave functions,
whereas the domain of a self adjoint WDW operator acting on tunneling (Vilenkin
type) wave function is trivially a single ray. One can define a positive
definite Hilbert space inner product using both kinds of wave functions.
If, on the other hand, we choose the scale factor of the Universe to
parametrize
time, both wave functions give rise to non trivial domains, and are independent
of any new parameters. However, positive norm physical states may be
constructed
using only Vilenkin type wave functions. Therefore, whenever we are free to
choose either matter fields or the scale factor as a time coordinate,
self-adjointness of the (spatial part of the) WDW operator dictates utterly
different physical Hilbert spaces corresponding to different physical
realities, and is therefore intimately related to the problem of time in
quantum gravity\cite{Kuchar}.

\par

In section 2 we consider a non-empty RW universe filled up with
matter in the form of scalar fields where we derive
(\ref{1.13})-(\ref{1.15}) explicitly. We show how the requirement for
self-adjointness of the (spatial part of the) WDW operator arises
and point its relation to the problem of time. In section 3 we discuss
self-adjoint extensions of one dimensional Schr\"odinger
operators whose potential terms are unbounded from below, but satisfy
(\ref{1.9}). In section 4 we apply such extensions to RW quantum cosmology and
discuss the Hartle-Hawking and Vilenkin proposals for the wave function of the
Universe in this context. Finally, we show in the appendix that the domain of
the self adjoint WDW operator (\ref{1.10}) acting on tunneling (Vilenkin type)
wave functions is a {\em single} ray.
\pagebreak

\vspace{0.5cm}
\section{Non-Empty RW Minisuperspaces}
\setcounter{equation}{0}
\vspace{0.4cm}

Consider a scalar field $\phi(\eta,\vec{x})$ which is coupled non-minimally
to gravity. It is governed by the action
  \be
S_{m} = - \frac{1}{2} \int_{M} d^{4}x \sqrt{-g} \left[
\partial_{\mu} \phi \partial^{\mu} \phi + \xi R \phi^{2} + V(\phi) \right]
- \frac{1}{2} \xi \int_{\partial M} d^{3}x \sqrt{g^{(3)}}
K \phi^{2}\,.
\label{2.11}
  \ee
where $\xi$ is the coupling constant.
In the homogeneous and isotropic case, we have $\phi = \phi(\eta)$,
and thus the total action (namely, the sum of (\ref{1.2}) and (\ref{2.11})) is
\cite{Hartle}
  \be
S_{tot} = S_{g} + S_{m} = \int d\eta N_{\perp}
\left[ - a \frac{\dot{a}^{2}}{N_{\perp}^{2}}
+ \frac{ a^{3 - 12\xi} {\dot{\chi}}^{2} }{ N_{\perp}^{2} }
+ U(a,\chi) \right]
\label{2.12}
  \ee
where $\chi \equiv \pi a^{6 \xi} \phi$ is the rescaled matter field and
  \be
U(a,\chi) = a - g^{2} a^{3} - \pi^{2} a^{3} V({\chi\over  \pi a^{6\xi}})
- 6\xi a^{1-12\xi} \chi^{2}\,.
\label{2.13}
  \ee
The corresponding Hamiltonian is
  \be
H_{WDW}\equiv H_{tot} = N_{\perp} \left[- \frac{1}{4a} P_{a}^{2} +
\frac{1}{4a^{3-12\xi}} P_{\chi}^{2} - U(a,\chi) \right]\,.
\label{2.14}
  \ee
In the coordinate representation $\hat{P}_{a} = -i
\frac{\partial}{\partial a}$ and $\hat{P}_{\chi} = -i
\frac{\partial}{\partial \chi}$, the WDW
equation reads
  \be
\left[ - \frac{1}{4} \frac{\partial^{2}}{\partial a^{2}}
+ \frac{1}{4} a^{12\xi -2} \frac{\partial^{2}}{\partial \chi^{2}}
+ a U(a,\chi) \right] \Psi(a,\chi) = 0 ~.
\label{2.15}
  \ee
We concentrate on the simple conformally invariant case where $V(\phi)=0$ and
$\xi = 1/6$, such that (\ref{2.14}) becomes
  \be
H_{WDW}\equiv H_{tot} = - \frac{N_{\perp}}{a} \left( \frac{1}{4} P_{a}^{2} -
\frac{1}{4} P_{\chi}^{2} + a^{2} - g^{2} a^{4} - \chi^{2}
\right) ~.
\label{2.16}
  \ee
The WDW equation is therefore
  \be
\left( - \frac{1}{4} \frac{\partial^{2}}{\partial a^{2}}
+ a^{2} - g^{2} a^{4} + \frac{1}{4}
\frac{\partial^{2}}{\partial \chi^{2}} - \chi^{2} \right)
\Psi(a,\chi) = 0 ~.
\label{2.17}
  \ee
We solve (\ref{2.17}) by separation of variables $\Psi(a,\chi)
= \psi_{a}(a) \psi_{\chi}(\chi)$, which results in the two coupled equations
  \begin{eqnarray}
\left( - \frac{1}{4} \frac{\partial^{2}}{\partial a^{2}}
+ a^{2} - g^{2} a^{4} \right) \psi_{a}(a) &=& E \psi_{a}(a)
\label {2.18} \\
\left( - \frac{1}{4} \frac{\partial^{2}}{\partial \chi^{2}}
+ \chi^{2} \right) \psi_{\chi}(\chi) &=&
E \psi_{\chi}(\chi)\,.
\label{2.19}
  \end{eqnarray}
Quantization requires gauge fixing, namely a definition of ``time". The freedom
left in making such a gauge choice leads to the ``problem of time" in quantum
cosmology \cite{Kuchar,Banks}.
It is strongly related to the definition of a physical
Hilbert space, and here is the place where considerations of self-adjointness
come into play\footnote{The discussion in \cite{Isham} is limited to
the case of zero cosmological constant for which there is no need for self
adjoint extensions of the WDW operator.}\cite{Isham}.
One has to introduce an inner product in the space of solutions
to the WDW equation (\ref{2.17}) in order to define the physical
Hilbert space. Defining the currents $J^{12}_{a,\chi} = i
\Psi_{1}^{*} \stackrel{\leftrightarrow}{\partial}_{a,\chi} \Psi_{2}$, we have
  \be
\partial_{a} J_{a}^{12} - \partial_{\chi} J_{\chi}^{12} = 0\,.
\label{current-eq}
  \ee
for {\it any} pair $\Psi_1 ,\Psi_2$ of solutions of (\ref{2.17})

In our simple minisuperspace model we can choose the time variable either as
the scale factor $a$ or as the matter field $\chi$ \cite{Isham,Vil,Kazama}.
If we choose the scale factor $a$ as our time, $t=a$, then
  \be
\lag\Psi_{1} | \Psi_{2}\rag_{(a)} = -i \int_{-\infty}^{\infty} d\chi
\left[ \Psi_{1}^{*}(a,\chi)
{\stackrel{\leftrightarrow}{\partial}}_{a} \Psi_{2}(a,\chi)
\right]_{|_{a=t=const.}}
\label{inner1}
  \ee
is a natural inner product to use in constructing the physical Hilbert space.

A well defined inner product among solutions of (\ref{2.17}) is necessarily
time independent in order that there be no conflict between time evolution of
physical states and the definition of Hilbert space at each time slice
$t={\rm constant}$. Deriving (\ref{inner1}) with respect to time we have
  \be
i\partial_{t} \lag\Psi_{1}|\Psi_{2}\rag_{(a)}
\equiv i\partial_{a} \lag\Psi_{1}|\Psi_{2}\rag_{(a)} = J_{\chi}^{12}(+\infty) -
J_{\chi}^{12}(-\infty)\,,
\label{con-eq1}
  \ee
where we have used (\ref{current-eq}) and integrated over $\chi$. Therefore,
time independence of (\ref{inner1}) implies
  \be
J_{\chi}^{12}(+\infty) - J_{\chi}^{12}(-\infty) = 0\,.
\label{j1}
  \ee
This condition holds automatically due to finiteness of
(\ref{inner1}) which also leads to a discrete spectrum of (\ref{2.19}). The
latter is the standard one dimensional harmonic oscillator spectrum, $E_{n} =
\omega (n + 1/2) = (n + 1/2)$, for which $J_{\chi}^{12}(+\infty) =
J_{\chi}^{12}(-\infty) = 0$ . The Schr\"odinger operator on the left hand side
of (\ref{2.19}) which is the spatial part of the WDW operator in
(\ref{2.16}) is clearly self-adjoint in this domain. Evidently, (\ref{j1})
sets no restrictions on the Schr\"odinger operator on the left hand side of
(\ref{2.18}) and therefore self-adjointness of the WDW operator
leads to no further consequences in this case. This implies, as we show in the
next two sections, that a non trivial domain of the self-adjoint WDW operator
includes either zero-current (zero-norm) wave functions or tunneling wave
functions.

Matters are utterly different when we choose $\chi$ (and not $a$) as time.
In this event, the roles of $\chi$ and $a$ interchange. One should integrate
(\ref{current-eq}) over $a$, and the inner product is therefore
  \be
\lag\Psi_{1} | \Psi_{2}\rag_{_{(\chi)}} = -i\int_0^{\infty} da \left[
\Psi_{1}^{*}(a,\chi)
\stackrel{\leftrightarrow}{\partial}_{\chi} \Psi_{2}(a,\chi)
\right]_{|_{\chi=t=const.}}.
\label{inner2}
  \ee
This inner product is time independent ($\chi$ independent)
{\em only} if
  \be
J_{a}^{12}(+\infty) - J_{a}^{12}(0) = 0
\label{hermit-con}
  \ee
for any pair of solutions $\Psi_{1,2}$ of (\ref{2.17}).
This condition is quite different from (\ref{j1}). Unlike the latter, it is
not satisfied automatically, because the potential energy in (\ref{2.18}) is
unbounded from below. Rather, it must be {\em imposed}, in order that the
Schr\"odinger operator on the left hand side of (\ref{2.18}) which is the
spatial part of the WDW operator in (\ref{2.16}) for this choice of
time parametrization, be well defined. Indeed, (\ref{hermit-con}) is precisely
the condition that the Hamiltonian on the left hand side of (\ref{2.18}) be
symmetric with respect to this inner product. This can be accomplished by
defining proper self-adjoint extensions of the Hamiltonian as we discuss in the
next two sections. Namely, we show there that (\ref{hermit-con}) implies that
zero current wave functions (similar in form to the Hartle-Hawking wave
function) give rise to a two parameter continuous family of non trivial domains
for the self adjoint WDW operator whose spectrum $\{E_n\}$ is {\em discrete}
and highly non linear in $n$. On the other hand, the domain of a self adjoint
WDW operator acting on tunneling (Vilenkin type) wave functions is a
{\em single} ray having a single eigenvalue.

Demanding that the Hamiltonian be self adjoint is equivalent to the requirement
of unique time evolution of quantum states. We see that the two different
choices of ``time" in this model lead to different kinds of
physical Hilbert spaces and therefore to two highly different physical
``realities". Thus, the requirement that the spatial part of the
WDW operator be self-adjoint is intimately related with the problem
of time in in quantum cosmology which manifests itself to its extreme in the
model discussed here.

We are not going to argue here which choice of time coordinate is superior to
the other. However, we feel that the choice $t=\chi$ has not received enough
attention (at least as far as mathematical aspects
are concerned), and we wish to fill this gap partially. On top of this,
note that implications of self-adjointness of the (spatial part of the)
WDW operator are less trivial here compared to those related with
the other possible choice.

In the non conformally invariant case, the WDW equation is
generally non-separable and therefore much more complicated.
However, in cases where $V(\chi)$ has a narrow deep minimum
at $\chi=\chi_s$ we can assume to a first approximation that $\chi$ does not
fluctuate far away from that minimum and thus replace $V(\chi)$ by its minimal
value $\rho_s = V(\chi_s)$. In this approximation one thus merely shifts
$g^{2}$ into $g_{eff}^{2} = g^{2} + \pi^2\rho_{s}$.

We close this section making some general remarks before turning to the formal
discussion of the proper self-adjoint extensions of (\ref{2.18}). Strictly
speaking, the scale factor $a$ in (\ref{RW}) is defined
on the ray $a\geq 0$. Therefore all Schr\"odinger operators involving $a$ must
be equipped with a suitable boundary condition at $a=0$ in order to make them
self-adjoint. Such a boundary condition is a necessary datum purely from the
Schr\"odinger theory point of view and for this reason we will impose such
a condition on wave functions below. However, which boundary condition at $a=0$
must be imposed on (\ref{1.10}) and (\ref{2.18}) in order to
describe quantum cosmology is a highly controversial issue. Indeed, unlike its
usefulness at large radii, the mini-superspace formalism we use in this paper
might break completely at extremely small radii of the Universe because of the
true singularity of (\ref{RW}) at the point $a=0$. Should this happen,
(\ref{1.10}) and (\ref{2.18}) will become useless at
$a \rightarrow 0$ as well as their solutions. The region $a >> 1/g$, on the
other hand, is certainly in the validity domain of the minisuperspace
quantization scheme.  We may thus trust the wave functions resulting from the
WDW equation only for large universe radii, as far as
cosmological interpretations are concerned.

As was discussed in the introduction, the fact that the time of flight
(\ref{1.9}) of a classical particle moving in the potential (\ref{1.8}) to
infinity is finite has a very important consequence. Namely, it effectively
turns the spectral problem involving the hamiltonian on the left hand side of
(\ref{1.10}) into a problem defined on a finite segment, despite the fact that
$0 \leq a < \infty$. This calls for an appropriate boundary condition on wave
functions at $a=\infty$ as well. The boundary condition imposed at $a=0$ must
be consistent with the one set at $a=\infty$. Thus, in principle, there is some
influence by the $a=0$ endpoint (where the minisuperspace formalism is
suspicious) on the asymptotic behavior of the wave function of the Universe as
$a \rightarrow \infty$ (where the minisuperspace formalism is surely valid).
We comment on this point in section 4.

\pagebreak

\vspace{0.5cm}
\section{Domain of the Self-Adjoint Hamiltonian  Defined on the Ray
$0 \leq x < \infty$}
\setcounter{equation}{0}

Consider a particle of mass $m$, moving in the one dimensional
potential
\be
V(x) = x^{2} - g^{2}x^{4}~,~ x \geq 0
\label{3.1}
\ee
having energy $ 0 \leq E \leq {1\over 4g^2}$. There are two classically allowed
regions for this range of energies, $x \leq x_{1}$ and $x \geq x_{2}$. Here
$x_{2}(E)>x_{1}(E)>0$ are the two classical turning points, namely, the two
positive real roots of $V(x)=E$.

A wave packet with energy distribution peaked at $E$ will move in
(\ref{3.1}) from $x_{i} \geq x_{2}$ to infinity in a {\em finite} period of
time
  \be
t_{E} = \int_{x_{i}}^{\infty} \frac{dx}{\sqrt{2m[E-V(x)]}}\,,
\label{3.2}
  \ee
raising the question what will happen to the wave packet as it ``hits" the
point
at infinity or equivalently, how is probability conserved in such a system.
Therefore, on account of (\ref{3.2}), unbounded motion in the potential
(\ref{3.1}) behaves in many respects as if it were bounded, and the point at
infinity appears as if it were the end point of a finite segment\cite{Farhi2}.

Probability conservation requires that
the Hamiltonian governing this system be self-adjoint. This requirement on the
domain of definition of the Hamiltonian is by no means trivial, since wave
functions in the potential (\ref{3.1}) have only power-like decay while their
first derivatives blow up at infinity, as can be most easily seen by writing
down the leading WKB approximation
\begin{eqnarray}
\Psi_E(x) & \sim & {1\over \left[E-V(x)\right]^{1\over4} } \left\{ C_{1}(E)
\, \mbox{sin} \left[\, \int
\limits_{x_2}^{x}\sqrt{2m(E-V(y))} dy - {\pi\over 4}\right] \right. \nonumber
\\
& & \left. + ~ C_{2}(E)\, \mbox{cos} \left[\, \int
\limits_{x_2}^{x}\sqrt{2m(E-V(y))} dy - {\pi\over 4}\right]\right\}
\label{3.3}
\end{eqnarray}
to a generic solution of the Schr\"odinger equation
\be
\left[-{1 \over 2m}{d^2\over dx^2} + V(x)\right]\Psi_E(x) = E\Psi_E(x)
\label{3.4}
\ee
in region $x > x_{2}$. Finiteness of $t_E$ in (\ref{3.2}) means that
(\ref{3.3}) is square integrable for {\em any} value of $E$, but this is not
true of its first derivative.
Due to the fact that $V(x\rightarrow\infty) \rightarrow -\infty$, local
De-Broglie wave lengths of the particle become extremely short very quickly as
it moves deeper into the classically allowed region rendering
the WKB approximation more and more accurate as $x\rightarrow\infty$
\footnote{Recall that this is also the region of scale factor $a=x$ values
where
minisuperspace analysis is most valid anyway.}. It is therefore enough to limit
our discussion to the framework of the WKB approximation\cite{Farhi2}. Such
asymptotic behavior of $\Psi(x)$ and $\Psi^{'}(x)$ as $x\rightarrow \infty$ is
in clear contrast with the exponential fall off of both bound state wave
functions and their first derivatives in case of potentials that are bounded
from below. In particular, it implies that there can be two square integrable
linearly independent solutions of (\ref{3.4}) sharing the same parameter $E$
because their constant Wronskian need not vanish. Thus, square integrability is
not sufficient to determine the spectrum, and what is needed is an explicit
boundary condition at infinity which should be treated as if it were really a
finite boundary point.

For {\sl any\/} two states $\Psi_1$ and
$\Psi_2$ in the domain of the self-adjoint Hamiltonian we must have
  \be
\lag\hat{H} \Psi_{1} | \Psi_{2}\rag\, = \,\lag\Psi_{1} | \hat{H} \Psi_{2}\rag
\,.
\label{3.5}
  \ee
Using the coordinate representation and integrating by parts, (\ref{3.5})
implies
  \be
\left[ \frac{d\Psi_{1}^{*}}{dx} \Psi_{2} -  \Psi_{1}^{*} \frac{d\Psi_{2}}{dx}
\right]^{+ \infty}_{0} = 0.
\label{3.6}
  \ee
This is precisely the consistency condition (\ref{hermit-con}) encountered
above.

Considering non trivial domains of the self adjoint operator $\hat H$, the
current $J^{12} = {i \over 2} (\Psi_{2} \partial_{x} \Psi^*_1 -
\Psi^*_1 \partial_{x} \Psi_2 )$
must vanish at the two boundary points, since otherwise
a particle reaching $x=\infty$ will have to reappear at $x=0$ or vice versa.
While such a {\em periodic} boundary condition is relevant for quantization of
a particle in a finite rigid box\cite{Farhi3}, it is clearly improper here
because in our case $V(0)=0$ while $V(\infty)=-\infty$. We back this
qualitative
argument by an explicit calculation presented in the appendix where we show
that if the current $J^{12}$ did not vanish, the domain of the self adjoint
operator $\hat H$ becomes trivial and collapses into a single ray.

Therefore, studying non trivial domains, (\ref{3.6}) may be replaced by the
stronger condition
  \be
\left[ \frac{d\Psi_{1}^{*}}{dx} \Psi_{2} -
\Psi_{1}^{*} \frac{d\Psi_{2}}{dx} \right] (x \rightarrow \infty)
= \left[ \frac{d\Psi_{1}^{*}}{dx} \Psi_{2} -
\Psi_{1}^{*} \frac{d\Psi_{2}}{dx} \right] (x = 0) = 0 .
\label{3.7}
  \ee

We show now that (\ref{3.7}) will {\em not} hold (at $x \rightarrow \infty$)
for generic $\Psi_{1} = \Psi_{E_{1}}$ and $\Psi_{2} = \Psi_{E_{2}};
\quad E_{1} \neq E_{2}$, {\em unless}
some special choice of the functions $C_{1}(E)$ and $C_{2}(E)$ in (\ref{3.3})
is made. To make this point clearer, it is useful to introduce the phase
$\phi_{\alpha}(E)$ defined by\cite{Farhi2}
  \be
\mbox{cot}\left( \phi_{\alpha}(E) \right) = - \frac{C_{2}(E)}{C_{1}(E)}
\label{3.8}
  \ee
where $\alpha$ is a parameter wave functions depend upon and will be
determined later. In terms of $\phi_{\alpha}(E)$, (\ref{3.3}) becomes
  \be
\Psi^{(\alpha)}_{E}(x > x_{2}) \sim -\frac{ C_1(E)}
{{\left[ E - V(x) \right]}^{\frac{1}{4}}
\mbox{sin}\phi_{\alpha}(E)} \mbox{cos} \left[ \int_{x_{2}(E)}^{x}
\sqrt{2m(E - V(y))} dy + \phi_{\alpha}(E) - \frac{\pi}{4} \right] .
\label{3.9}
  \ee
The cosine in (\ref{3.9}) oscillates very rapidly as $x \rightarrow \infty$
and therefore (\ref{3.7}) will not be met (for $E_{1}\neq E_{2}$) unless the
argument of the cosine becomes independent of $E$ as $x \rightarrow \infty$,
namely,
  \be
\frac{\partial}{\partial E} \left[ \int_{x_{2}(E)}^{x}
\sqrt{2m(E-V(y))} dy + \phi_{\alpha}(E) \right] = 0\quad,\quad x\rightarrow
\infty\,.
\label{3.10}
  \ee
In order to solve (\ref{3.10}) we need to specify an initial condition in $E$,
this is how the parameter $\alpha$ gets in. One can choose
  \be
\phi_{\alpha}(E=\alpha) = 0
\label{3.11}
  \ee
and the solution of (\ref{3.10}) (subjected to (\ref{3.11})) is \cite{Farhi2}
  \be
\phi_{\alpha}(E) = \sqrt{2m}
\left\{ - \int_{x_2(E)}^{\infty} \left[
\sqrt{E - V(y)} - \sqrt{\alpha - V(y)} \right] dy +
\int_{x_{2}(\alpha)}^{x_{2}(E)} \sqrt{\alpha - V(y)} dy \right\}
\label{3.12}
  \ee
where $x_{2}(\alpha)$ is the largest root of $V(x) = \alpha$. Substituting
the specific potential (\ref{3.1}) into (\ref{3.10}) and integrating over $E$
we obtain
\begin{eqnarray}
\phi_\alpha (E) = \left\{\begin{array}{ccc} {1\over g^2}
\int_{z(\alpha)}^{z(E)} \zeta^2 {\cal K}({\zeta^2-1\over 2\zeta^2}) d\zeta &
&;~E\leq 0 \\ & & \\{1\over g^2}\left[C_\alpha + \int_1^{z(E)}
{\sqrt{2}\zeta^3 \over \sqrt{1+\zeta^2}} {\cal K}({1-\zeta^2\over 1+\zeta^2})
d\zeta \right] & &;~E\geq 0 \end{array}\right.
\label{phi}
\end{eqnarray}
where $z(E)=(1-4Eg^2)^{1\over 4}$, $C_\alpha=\int_{z(\alpha)}^1
\zeta^2 {\cal K}({\zeta^2-1\over 2\zeta^2}) d\zeta $ and  ${\cal K}(m)$
(${\cal E}(m)$) is a complete
elliptic integral of the first (second) kind (we use ${\cal E}(m)$ below). Here
we have also assumed $\alpha < 0$ so that $C_{\alpha}<0$ as well .

{}From (\ref{3.12}) we see that for $x \rightarrow \infty$ (\ref{3.9})
approaches
  \be
\Psi^{(\alpha)}_{E}(x \rightarrow \infty) =
- \frac{ C_{1}}{{\left[ E-V(x) \right]}^{\frac{1}{4}}
\mbox{sin}\phi_{\alpha}(E)} \mbox{cos} \left[
\int_{x_{2}(\alpha)}^{x}
\sqrt{2m(\alpha - V(y))} dy - \frac{\pi}{4} \right]
\label{3.13}
  \ee
where the argument of the cosine is indeed $E$ independent. It is clear
now that (\ref{3.7}) will hold at $x \rightarrow \infty$ for any pair of
functions $\Psi^{(\alpha)}_{E_{1}}\,,\,\Psi^{(\alpha)}_{E_{2}}$ of the form
given by (\ref{3.9}) and (\ref{3.12}). These functions form a family of
solutions of the Schr\"{o}dinger equation (\ref{3.4}) parametrized by the
single (real) parameter $\alpha$. These are not {\em eigenstates} of the
Hamiltonian, as we have not imposed (\ref{3.7}) at $x=0$ yet. To this end
we note that a necessary and a sufficient condition for vanishing of the
probability current at $x=0$ is
  \be
\frac{\Psi'_{E}(x=0)}{\Psi_{E}(x=0)} = \beta
\label{3.14}
  \ee
where $\beta$ is a fixed real
number\cite{von-Neumann,Reed-Simon,Farhi1,Farhi3}.

In a similar manner to (\ref{3.9}) we write the WKB solution of (\ref{3.4}) as
  \be
\Psi^{(\beta)}_{E}(x < x_{1}) = \frac{C_{3}(E)}
{\left[ E - V(x) \right]^{\frac{1}{4}}
\mbox{sin} \xi_{\beta}(E)}
\mbox{cos} \left[ \int_{x}^{x_{1}(E)}
\sqrt{2m(E - V(y))} dy + \xi_{\beta}(E) \right]
\label{3.15}
  \ee
and (\ref{3.14}) fixes
  \be
\xi_{\beta}(E) = \mbox{arctan}\left( \frac{\beta}{\sqrt{2mE}} \right)
- \int_{0}^{x_{1}(E)} \sqrt{2m(E-V(y))} dy\,.
\label{3.16}
  \ee

Eigenstates of (\ref{3.4}) are obtained by matching (\ref{3.9}) (subjected to
(\ref{3.12})) and (\ref{3.15})-(\ref{3.16}). Using the ordinary WKB matching
conditions we obtain the corresponding ``Bohr-Sommerfeld" quantization
condition on $E$, namely,
  \be
4\,\mbox{tan} \phi_{\alpha}(E)
= e^{-2\Delta(E)} \mbox{tan} \left[ \frac{\pi}{4} - \xi_{\beta}(E)
\right] ,
\label{3.17}
  \ee
where
  \beqra
\Delta(E) & = &\int_{x_{1}(E)}^{x_{2}(E)}\sqrt{2m|E - V(y)|} dy \nonumber\\
&{}&\nonumber\\
& = &{\sqrt{2m} x_2(E)\over 3g}\left[ {\cal E}\left(1- {x_1(E)^2\over
x_2(E)^2}\right) - 2 g^2 x_1^2(E) {\cal K}\!\left(1- {x_1(E)^2\over
x_2(E)^2}\right) \right]\,.
\label{delta}
  \eeqra

So far we have concentrated on energy range $ 0 \leq E \leq {1\over 4g^2}$.
Our discussion may be extended in a straightforward manner to the complementary
energy ranges $E>{1\over 4g^2}$ and $E\leq 0$ as well. For example, the
quantization condition corresponding to $E>{1\over 4g^2}$ reads
  \be
\mbox{tan} \left( \int_{0}^{\infty} \left( \sqrt{2m(E-V(y))}
- \sqrt{2m(\alpha - V(y))} \right) dy \right) =
-\frac{\beta}{\sqrt{2mE}} .
\label{3.18}
  \ee

Note the explicit dependence of energy eigenvalues and eigenfunctions
upon $\alpha$ and $\beta$. The spectrum is therefore a two parameter family
$\{ \Psi^{(\alpha,\beta)}_E(x) \} $, parametrized by $\alpha$ and $\beta$ as
it should be in this case of separeted boundary conditions
according to the general theory of self-adjoint
extensions\cite{von-Neumann,Reed-Simon,Farhi1,Farhi3}. Due to (\ref{3.2}) the
point at infinity appears as if it were a finite endpoint and the whole
quantum system behaves as if it were defined on a finite segment, where
$\alpha$ and $\beta$ parametrize boundary conditions at the two endpoints.
The set of functions $\{ \Psi^{(\alpha,\beta)}_E(x) \} $ spans the domain of
the self-adjoint Hamiltonian, namely, the space of all square integrable
functions that satisfy (\ref{3.7}).

Note that the WKB density of states
   \be
\rho_{_{WKB}}(E)= {m\over \pi}\int_0^{\infty} dx\,{{\cal \theta} [E-V(x)]\over
\sqrt{2 m [ E - V(x)]}}
\label{density}
    \ee
which is proportional to (\ref{3.2}) is finite (as long as $E\neq
{1\over 4g^2}$)\footnote{For $E\rightarrow {1\over 4g^2}$ (\ref{density})
diverges logarithmically in $|E- {1\over 4g^2}|$.}. The spectrum must be
therefore discrete for {\em any} $E\neq {1\over 4g^2}$ (with an accumulation
point at ${1\over 4g^2}$). It is moreover bounded neither from below, nor from
above and contains therefore an infinite amount of discrete states. Thus, the
domain ${\cal D}^{\alpha,\beta}$ of the self-adjoint
Hamiltonian is the set of all discrete linear combinations of the form
  \be
\Psi^{(\alpha,\beta)}(x) = \sum_{n} c_{n} \Psi_{E_{n}}^{(\alpha,\beta)}(x) ,
\label{3.19}
  \ee
where $\sum_{n} |c_{n}|^{2} < \infty$ which yields an {\em infinite}
dimensional
Hilbert space.
\pagebreak

\section{Quantum Cosmological Implications}
\setcounter{equation}{0}

The most important observation made in section 2 (as far as the simple
cosmological model discussed there is concerned) is that time independence of
the inner product in the definition of the physical Hilbert space leads to
the requirement that the spatial part of the WDW operator, which is
a one dimensional Schr\"odinger operator, be self adjoint. We saw in that
section that one can choose time parametrization either in terms
of the matter field $\chi$ or in terms of the scale factor $a$ of the Universe
and that these two choices of time parametrization lead to different physical
realities. In our view this is an extreme manifestation of the problem of
time in quantum cosmology.

Our discussion in the previous two sections makes it clear that one may trace
this discrepancy of physical realities to the fact that the two choices of time
parametrization lead to two utterly different physical Hilbert spaces. The
reason for this difference stems directly from the requirement that the
(spatial part of the) WDW operator be self-adjoint. In this section
we sharpen this distinction and investigate its cosmological implications in
more detail.

Let us first concentrate on parametrization of time in terms of the scale
factor
$a$. From the discussion following (\ref{inner1}) it is clear that a generic
physical state has the form
  \be
\Psi(a,\chi)=\sum_{n\geq 0}c_n\psi_{an}(a)\psi_{\chi n}(\chi)
  \label{4.1}
  \ee
where $c_n$ are complex constants, $\psi_{\chi n}(\chi)$ is the normalized
harmonic oscillator eigenstate corresponding to eigenvalue $E_n = (n+ {1\over
2})$ and $\psi_{an}(a)$ is a normalized solution of (\ref{2.18}) with parameter
$E=E_n$\,\footnote{
Recall from our discussion in section 3 that these solutions are normalizable
because of the finiteness of (\ref{3.2}), as long as $E\neq {1\over 4g^2}$.}
so that matter excitations are seemingly those of a free field.
Because of orthonormality of the $\psi_{\chi n}(\chi)$ the inner product
(\ref{inner1}) of any two such states is simply
  \be
\lag\Psi^{(1)} | \Psi^{(2)}\rag_{(a)} = \sum_{n\geq 0} j_n c_n^{(2)^*}
c_n^{(1)}
\label{4.2}
  \ee
where $j_n =-i[\psi_{an}^*(a)\psi_{an}(a)'-\psi_{an}^*(a)'\psi_{an}(a)]$
is the ``Schr\"odinger" current carried by $\psi_{an}(a)$. This current is
clearly $a$ independent, making (\ref{4.2}) time independent in accordance with
(\ref{con-eq1}). The ``Schr\"odinger" current carried by
$\psi_{an}(a)$ must be positive in order that (\ref{4.2}) be positive definite.
This positivity condition becomes simply
  \be
|B_1(E)|^2-|B_2(E)|^2 >0
\label{4.3}
  \ee
where $B_1$ and $B_2$ are the amplitudes of the outgoing and of the incoming
waves in the generic solution of (\ref{2.18}) to the right of the largest
turning point $a_2$ ($V(a_2)=0$)\footnote{Recall that here $E=E_n>0$. For all
practical uses we assume further that $E<{1\over 4g^2}$ so that $a_2>0$.}
\begin{eqnarray}
\psi_{aE} (a > a_2) &=& { 1\over \left[ E - V(a) \right]^{1\over 4}}
\left\{ B_1(E)\, \mbox{exp} \left[ i\int_{a_2 (E)}^{a}
\sqrt{4(E - V(y))} dy \right] +\right.
\nonumber\\
&&{}\nonumber\\
&&\left. B_2(E)\, \mbox{exp} \left[ -i
\int_{a_2 (E)}^{a} \sqrt{4(E - V(y))} dy \right]\right\}\,.
\label{4.4}
\end{eqnarray}
For $B_2=0$, (\ref{4.4}) gives pure expansion modes of the metric. The lowest
mode (corresponding to $n=0$) is the original
wave function suggested by Vilenkin\cite{Vilenkin}.
Wave functions of the form (\ref{4.4}) subjected to (\ref{4.3}) may be referred
to as ``generalized tunneling wave functions" \cite{Vilen2}.
Note further that if we have $|B_1(E)|=|B_2(E)|$ for all
$\psi_{an}(a)$ in (\ref{4.4}) the latter describes standing wave modes of the
form
  \be
\psi_{aE} (a > a_2) = { B(E)\over \left[ E - V(a) \right]^{1\over 4}}
\left\{ \mbox{cos}  \left[ \int_{a_2 (E)}^{a}
\sqrt{4(E - V(y))} dy +\phi (E) - {\pi \over 4}\right] \right\}
\label{4.5}
  \ee
carrying no current, that is $j_n=0$ for all $n$. The lowest of these modes
($n=0, \phi(E)=0$) is the wave function suggested by Hartle and
Hawking\cite{Hart-Haw,Hawking,Hartle}.
Following (\ref{4.2}) one cannot associate probabilistic
interpretation with such wave functions because all physical states in
(\ref{4.1}) become zero norm states.

We turn now to the case where time is parametrized by the matter
field $\chi=\pi a \phi$. Here the spatial part of the WDW operator
is the Schr\"odinger operator in (\ref{2.18}), which requires
non-trivial self-adjoint extensions as we saw in the previous section. It is
shown in the appendix that in this case the domain of the self adjoint WDW
operator acting on tunneling (Vilenkin type) wave functions is trivial, namely,
a {\em single} ray and the spectrum shrinks to a single point $E=E_0$. It is
very interesting that the seemingly innocent requirement that the WDW operator
be self adjoint is so powerful that it allows only a single physical state to
exist. We stress that such a trivial domain occurs {\em only} for tunneling
type wave functions. This may have the far reaching cosmological implication
as the mechanism that selects the {\em unique} wave function of the
Universe\footnote{Note that the requirement of self adjointness by it self
cannot fix $E_0$ which will hopefully be determined by the exact short distance
Hamiltonian of quantum gravity.}.

The zero current wave functions span non-trivial domains of this self
adjoint WDW operator. In the classically allowed region $a>a_2$ (which is the
region where the minisuperspace approach is most reliable in the first place)
the WKB approximation to the eigenfunctions spanning the non trivial domain are
found from (\ref{3.9}) as
  \be
\psi_{an}^{(\alpha,\beta)}(a > a_{2}) = { C(E_n)\over
\left[ E_n - V(x) \right]^{1 \over 4}}
\left\{\mbox{cos} \left[ \int_{a_{2}(E_n)}^{a}
\sqrt{4(E - V(y))} dy + \phi_{\alpha}(E_n) - \frac{\pi}{4} \right]\right\}
\label{4.6}
  \ee
where $\phi_{\alpha}(E) $ is given by (\ref{3.12}) and $E_n\equiv
E_n^{(\alpha,\beta)}$ are solutions of (\ref{3.17}). The modes given by
(\ref{4.6}) that appear in physical cosmological states must have non-negative
energies $E_n \geq 0$ otherwise the energy condition for matter fields will be
violated. We note here that for most practical purposes only energies in the
range $0\leq E_n \leq {1\over 4g^2}$ are important.

Clearly there is a continuous family of such domains parametrized by the two
real continuous variables $\alpha$ and $\beta$. The functions (\ref{4.6}) are
standing wave modes of the geometry which are similar in form to the
Hartle-Hawking wave function, but do not correspond to the ``no-boundary"
proposal for $E>0$.

A generic physical state (corresponding to time parametrization by the
matter field $\chi=\pi a \phi$) has the form
  \be
\Psi^{(\alpha,\beta)}(a,\chi)=\sum_{n\geq 0}\psi^{(\alpha,\beta)}_{an}(a)
\psi_{\chi n}(\chi)
  \label{4.7}
  \ee
where now $\psi_{\chi n}(\chi)$ are solutions of (\ref{2.19}) with parameter
$E=E_n$. The functions $\psi_{\chi n}(\chi)$ are therefore generic parabolic
cylinder functions\cite{Abram-Steg,Bender} rather than harmonic oscillator
eigenstates because the $\{E_n\}$ in (\ref{3.17}) are obviously not harmonic
oscillator eigenvalues. It is convenient to express the general solution of the
parabolic cylinder equation (\ref{2.19}) in terms of the two linearly
independent functions $U_n$ and $V_n$ as\cite{Abram-Steg}
  \be
\psi_{\chi n}(\chi)=b_n U_n\left(2\chi\right) +
c_n V_n\left(2\chi\right) ,
  \ee
where
  \beqra
U_n(z) &=&
D_{E_n-{1\over 2}}(z)
{}~~~~~ \mbox{and} \nonumber\\
&&{}\nonumber\\
V_n(z) &=&
{\Gamma({1\over 2}-E_n)\over \pi} \left[
 D_{E_n-{1\over 2}}(-z) -
\mbox{sin}(\pi E_n) D_{E_n-{1\over 2}}(z)\right]\,.\nonumber
\label{4.8}
  \eeqra
Here $D_{\lambda}(z)$ are Whittaker functions, and
$\{b_n\,,\,c_n\}$ are complex constants.

The asymptotic behavior of the particular combinations of Whittaker functions
appearing in (\ref{4.8}) as $z\rightarrow\pm\infty$
is\cite{Abram-Steg,Bender} $U(z) \sim
e^{\mp z^2}|z|^{\pm E - {1\over 2}}$
and $V(z) \sim e^{\pm z^2}|z|^{\mp E - {1\over 2}}$,
and their Wronskian is $\sqrt{{2\over\pi}}$.

Because of orthonormality of the $\psi^{(\alpha,\beta)}_{an}(a)$ the ($\chi$
independent) inner product (\ref{inner2}) of any two states of the form
(\ref{4.7}) becomes
  \be
\lag\Psi^{(1)} | \Psi^{(2)}\rag_{(\chi)} =-i\sqrt{{8\over\pi}} \sum_{n\geq
0} (b_n^{(1)*}c_n^{(2)}-c_n^{(1)*}b_n^{(2)})\,.
\label{4.9}
  \ee
In order that physical states have positive norm we must impose
$Im(b_n^*c_n)>0$, a possible choice being $c_n=ib_n$\footnote{Note that using
harmonic oscillator eigenstates for the $\psi_{\chi n}(\chi)$ makes all states
in (\ref{4.7}) zero norm states in the same way that using (\ref{4.5}) for the
$\psi_{an}(a)$ renders all states (\ref{4.1}) zero norm states.}.

The {\em physical} Hilbert space inner product is
(\ref{4.9}) and it is finite provided $\{b_n, c_n\}$ are properly restricted,
for example, by demanding that $\sum_n ( |b_n|^2+|c_n|^2)<\infty$.
Following (\ref{3.19}), note however, that in order that (\ref{4.7}) as a whole
be in the Hilbert space of the self adjoint operator (\ref{2.18}) we must have
$N(\chi)=\sum_{n} |\psi_{\chi n}(\chi)|^{2} < \infty$. This condition will
hold generally for finite values of $\chi$ provided (\ref{4.9}) is finite.
However, from the asymptotic behavior of $U_n$ and $V_n$ we know that $N(\chi)$
blows up like $e^{\chi^2}$ as
$|\chi|\rightarrow\infty$ throwing (\ref{4.7}) out of the domain of
(\ref{2.18}) at $|\chi|=\infty$, eventhough (\ref{4.9}) remains finite. The
limit $|\chi|\rightarrow\infty$ is easily attainable for any finite value of
the scalar field $\phi$ as the scale factor $a$ blows up to infinity because
$\chi=\pi a \phi$. This apparent difficulty deserves further investigation,
but we are not going to do so in this paper. We only mention that this problem
can be avoided in the single tunneling wave function domain, because the single
energy $E_0$ can be always chosen as one of the harmonic oscillator eigenvalues
with the corresponding square integrable $\psi_{\chi E_0}(\chi)$.

We see that a single (and relatively simple) physical system, namely,
a conformal scalar field coupled to gravity, is described by two different and
incompatible physical Hilbert spaces, corresponding to the two different
possible time parametrizations. This implies that two highly different physical
``realities" correspond to the same field theory coupled to gravity
which in our opinion is an extreme manifestation of the problem of time
in quantum gravity\footnote{Using the terminology of
Kucha\v{r}\cite{Kuchar}, this is the ``Multiple Choice Problem".}.

In the model discussed above, time parametrization by the matter field $\chi$
forces us to use the gravitational wave functions
$\psi_{an}^{(\alpha,\beta)}(a)$ in (\ref{4.6}) (for $a>a_2$) in
constructing physical states (\ref{4.7}) in the non trivial domain. This means
that there is a {\em continuum} of distinct physical Hilbert spaces that are
parametrized by the two real variables $\alpha$ and $\beta$ which are the
domains for the non trivial self adjoint extensions of (\ref{2.18}). Thus,
specifying a physical state requires fixing $\alpha$ and $\beta$ first.

The most urging physical question that arises concerns the cosmological
interpretation of $\alpha$ and $\beta$: is there really a continuum of distinct
WDW operators or are these parameters fixed somehow by a yet unknown
dynamical mechanism associated with quantum gravity\footnote{As happens,
for example, with the $\theta$ angle in the Standard Model.}? If there is
such a continuum of WDW operators, which of them corresponds to the
``real" Universe?

To partially answer these questions, recall that the wave
function of the Universe is a unique solution of the WDW equation that must be
singled out of all the other mathematically possible solutions whose general
WKB form is (\ref{4.7}). One then generally argues\cite{Banks} that the
distinguished semiclassical solution which is the wave function of the Universe
will be picked up by matching it to a yet unknown solution of the WDW equation
corresponding to the {\em exact} theory of quantum gravity that governs
small geometries. In the case discussed here more is required, namely,
that such a matching condition will teach us something about $\alpha$ and
$\beta$ as well. It might be that the parameter $\beta$ is an artifact of our
minisuperspace approach, because it defines the boundary condition (\ref{3.14})
at $a=0$ where the RW minisuperspace approach as a whole probably breaks down.
Thus, it is plausible that if we knew the exact WDW operator for small
geometries the ambiguity associated with $\beta$ would be lifted either by
showing that $\beta$ is indeed a minisuperspace artifact, or by fixing its
value if it has anything to do with the exact Hamiltonian. Unfortunately, we
cannot see how such considerations apply as far as the parameter $\alpha$ is
concerned. Indeed, $\alpha$ specifies the boundary condition at
$a\rightarrow\infty$ where the semiclassical approximation is perfectly valid,
and moreover, it is completely independent of $\beta$. It seems to us highly
unlikely that small geometry quantum gravity effects will be able to remove the
ambiguity in the WDW operator and the wave function of the Universe associated
with $\alpha$.

As a matter of fact, $\alpha$ has the following interesting cosmological
interpretation: The original Hartle-Hawking wave function of the Universe
\cite{Hart-Haw,Hawking,Hartle} $\Psi_{HH}(a > a_{2}) \sim (|V(a)|)^{-1/4}
\mbox{cos} \left( \int_{a_{2}(E)}^{a} \sqrt{4|V(y)|}dy - \pi/4
 \right)$ is obtained from the semiclassical approximation to the Euclidean
path
integral based on the ``no boundary" proposal\cite{Hall-Hart}. It is governed
by one of the two Euclidean solutions to the classical equations of motion
having the larger action. It has been shown, however, in \cite{Louko} that the
contribution to the path integral from the other classical solution is relevant
as well. It follows than, that if one considers contributions to the path
integral weighed by {\em real} coefficients, one obtains a generalization of
the Hartle-Hawking wave function given by (\ref{4.6}) evaluated at $E=0$.
Thus, $\mbox{cos}\phi_{\alpha}(0)$ weighs the contribution of the larger action
saddle point and $\mbox{sin} \phi_{\alpha}(0)$ weighs the other one.

The ``no boundary" proposal of Hartle and Hawking points at the special
role played by zero energy $E=0$. It corresponds to regular Euclidean
geometries, whereas singular Lorentzian geometries occur for $E>0$.
Note further that $\beta$, whose interpretation is problematic from
cosmological point of view, disappears from (\ref{3.16}) and (\ref{3.17})
at $E=0$. It is therefore interesting to study the consequences of including
$E=0$ in the {\em discrete} spectrum of (\ref{3.17}). Using (\ref{phi}),
(\ref{3.16}) and (\ref{delta}) at $E=0$ and demanding that (\ref{3.17}) be
satisfied at $E=0$ as well, we obtain the following functional relation
between $g^2$ and $\alpha$

  \be
\mbox{tan} \left({C_{\alpha}\over g^2}\right) = -\,{1\over 4}
\mbox{exp} \left( - {2\over 3 g^2} \right) \,.
\label{4.11}
 \ee
where $C_{\alpha}$ was defined following (\ref{phi}). Thus, for a given value
of $\alpha$ (\ref{4.11}) {\em quantizes} the cosmological constant
$\Lambda=3g^2$. Note further that (\ref{4.11}) has a maximal positive solution
$g^2_{max}$ which sets an upper bound for the cosmological constant. For
realistic values of the cosmological constant $g^2 <<1$ Eq.(\ref{4.11})
reduces to
  \be
\Lambda_n = {3 |C_{\alpha}|\over n \pi }\quad\quad,\quad\quad
n>>{|C_{\alpha}|\over \pi}\,.
\label{4.12}
  \ee

We close this section by making some comments regarding possible extension of
our work. For simplicity, we have focused our discussion on RW cosmologies that
are isotropic and homogeneous. Generalization of our work to more complicated
cosmologies (presumably with positive cosmological constant) is clearly
required. Moreover, implications of uniqueness of the tunneling wave
function as the wave function of the Universe in case of time parametrization
by matter fields to inflationary models should be clarified. Work on these
points is in progress.

\vspace{1cm}
{\bf Acknowledgments} \\
We would like to thank B. Allen, J. Avron, J. Friedman and especially
W. Fischler, J. Louko, Y. Ne'eman, L. Parker, B. Simon and A. Vilenkin for
helpful discussions and correspondence. We are also indebted to N. A. Lemos
for pointing out an error in a preliminary version of this paper.

\newpage

\appendix

\section{Appendix}
\setcounter{equation}{0}

Consider the self adjoint Hamiltonian $H = {1\over 2m} P^2 + V(x)\,;\,x\geq 0$
and its
domain ${\cal D}$. We assume that $V(x)$ is unbounded from below such that
(1.9) holds. For simplicity we also assume that $V(0)=0$ is a local
extremum of $V(x)$\footnote{Note further that $V''(0)> 0 $ is a necessary
condition for $E\sim 0$ to be in the spectrum.}. We show in this appendix
(within the framework of the WKB approximation) that if ${\cal D}$ consists
purely of right moving (or of left moving) waves, it is trivial, namely, a
single ray. We also present indications that this is the case for generalized
tunneling wave functions as well.

We first consider right moving waves, but the results are the same for left
moving waves.
Let $x_{>}(E)$ be the largest turning point of a
classical particle with energy $E$ moving in a
general potential $V(x)$ satisfying (1.9).
A right moving WKB solution to the right of
$x_{>}(E)$, is
  \be
\Psi^{^{(WKB)}}_{E}(x>x_{_{>}}) = {A_{_{E}}\over [2m(E-V(x))]^{1/4}}
\,\mbox{exp} \left( i \int_{x_{_{>}}(E)}^x
\sqrt{2m(E-V(y))}dy \right) ,
\label{A1}
  \ee
where $A_{_{E}}$ is a complex amplitude.

For energies above the maximum $V_{max}$ of $V(x)$ (that is $E > {1\over 4g^2}$
in the case of (3.1)) there are no turning points and (\ref{A1})
is the wave function for all $x \geq 0$ so that $x_{_{>}}=0$.

Consider the current
  \be
J^{12}(x) = {i\over 2}(
\Psi_2 \partial_{x} \Psi^*_1 - \Psi^*_1 \partial_x \Psi_2)(x)\,.
\label{Acurrent}
  \ee
At $x=0$ we have
  \be
\left( J^{12}(0) \right)_{_{E>V_{max}}} =
\frac{1}{2} A_{_{E_{1}}}^* A_{_{E_{2}}} \left[
\left( {E_{1}\over E_{2}} \right)^{1/4} +
\left( {E_{2}\over E_{1}} \right)^{1/4} \right]
\label{A2}
  \ee
while for $x \rightarrow \infty$ the current becomes
  \be
J^{12}(x \rightarrow\infty )
\longrightarrow  A^*_{_{E_{1}}}A_{_{E_{2}}} e^{-i\delta_{12}(\infty)}\,.
\label{A3}
  \ee
Here $\delta_{12} (x) = \int_{x_{_{>}}(E_{2})}^{x} \sqrt{2m(E_{2}-V(y))}
dy - \int_{x_{_{>}}(E_{1})}^{x} \sqrt{2m(E_{1}-V(y))} dy $.
{}From (\ref{A2}) and (\ref{A3}) we see that $J^{12}(0)=J^{12}(\infty)$
{\em if and only if} $E_{1}=E_{2}$, so ${\cal D}$ contains at most only a
{\em single} eigenstate in the energy range $E>V_{max}$.

For energies below the lowest local minimum $V_{min}$ of $V(x)$ (that is,
$E<0$ in the case of (3.1)) there is only one turning point, $x_{0}(E)$, and
the current at $x=0$ is
  \begin{eqnarray}
(J^{12}(x=0))_{_{E<V_{min}}} &=& {1\over 2} A^*_{_{E_{1}}} A_{_{E_{2}}}
\left\{ \mbox{cosh}(\xi_{1}-\xi_{2}) \left[ \left( {E_{1}\over E_{2}}
\right)^{1/4} + \left( {E_{2}\over E_{1}} \right)^{1/4} \right]
\right. \\
& & \left. - i \left( {1\over 4} e^{-(\xi_{1}+\xi_{2})} - e^{\xi_{1}+\xi_{2}}
\right) \left[ \left( {E_{2}\over E_{1}} \right)^{1/4} -
\left( {E_{1}\over E_{2}} \right)^{1/4} \right] \right\},
  \nonumber
  \end{eqnarray}
where $\xi_{i} = \int_{0}^{x_{0}(E_{i})} \sqrt{2m(V(y)-E_{i})} dy$.
The norm squared of the current at $x=0$ is
  \begin{eqnarray}
|J^{12}(0)|^2_{_{E<V_{min}}} &=& {1\over 4} |A_{_{E_{1}}}^*A_{_{E_{2}}}|^2
\left\{ \mbox{cosh}^2 (\xi_{1} - \xi_{2})
\left( 2 + \sqrt{{E_{1}\over E_{2}}} +
\sqrt{{E_{2}\over E_{1}}} \right) \right.
\nonumber \\
& & \left. + \left( {1\over 4} e^{-(\xi_{1}+\xi_{2})} - e^{\xi_{1}+\xi_{2}}
\right)^2 \left( \sqrt{{E_{1}\over E_{2}}} +
\sqrt{{E_{2}\over E_{1}}} -2 \right) \right\},
\label{A4}
  \end{eqnarray}
while from (\ref{A3}) the norm squared at $x \rightarrow \infty$ is
  \be
|J^{12}(x \rightarrow \infty)|^2 = |A^*_{_{E_{1}}} A_{_{E_{2}}}|^2 .
\label{A5}
  \ee
The last two expressions are equal if and only if $E_1=E_2$ and thus also in
this energy range ${\cal D}$ contains at most a single eigenstate.

For energies in the range $V_{min}<E<V_{max}$ there are
more turning points and $J^{12}(0)$ is sensitive to the details of the
potential, but one can show that the domain ${\cal D}$ is trivial as well.
For concreteness let us consider the potential (3.1). In the energy range
$0 < E < {1\over 4g^{2}}$ there are two (positive) turning points,
$0<x_{_{<}}<x_{_{>}}$.
Using the WKB matching conditions, the wave function
to the left of $x_{_{<}}(E)$ which corresponds to (\ref{A1}) is
  \be
\Psi^{^{(WKB)}}_{E}(x<x_{_{<}}) = {-A_{_{E}} e^{i\pi/4} \over
[2m(E-V(x))]^{1/4}} \left\{
{e^{-\Delta(E)}\over 2} \mbox{sin} [\phi_{_{E}}(x)] +
2i e^{\Delta(E)} \mbox{cos} [\phi_{_{E}}(x)] \right\},
\label{A6}
  \ee
where $\phi_{_{E}}(x) = \int_{x}^{x_{_{<}}(E)} \sqrt{2m(E-V(y))}dy
- {\pi\over 4}$ and $\Delta(E)$ is given by ({\ref{delta}).
The current at $x=0$ becomes
  \begin{eqnarray}
J^{12}(0) &=& {1\over 2} A^*_{_{E_{1}}}A_{_{E_{2}}} \left\{
\left[ (c_{12} + c_{12}^{-1}) \mbox{cos}\phi_{1} \mbox{cos}\phi_2
\right. \right. \nonumber \\
& & + \left. (d_{12} + d_{12}^{-1}) \mbox{sin}\phi_1
\mbox{sin}\phi_2 \right] \nonumber\\
& & -i \left. \left[ (a_{12} + a_{12}^{-1}) \mbox{cos}\phi_{1} \mbox{sin}
\phi_{2} - (b_{12} + b_{12}^{-1}) \mbox{sin}\phi_{1} \mbox{cos}\phi_{2}
\right]
\right\},
\label{A7}
  \end{eqnarray}
where
  \begin{eqnarray}
a_{12} = {1\over 4} \left( {E_{1}\over E_{2}} \right)^{1/4}
e^{-(\Delta(E_{1})+\Delta(E_{2}))}
  ~~~ & , & ~~~~
b_{12} = {1\over 4} \left( {E_{2}\over E_{1}} \right)^{1/4}
e^{-(\Delta(E_{1})+\Delta(E_{2}))}   \\
c_{12} = \left( {E_{1}\over E_{2}} \right)^{1/4}
 e^{-(\Delta(E_{1})-\Delta(E_{2}))}
  ~~~~~ & , & ~~~~~
d_{12} = \left( {E_{2}\over E_{1}} \right)^{1/4}
e^{-(\Delta(E_{1})-\Delta(E_{2}))} , \nonumber
\label{abcd}
  \end{eqnarray}
and
  \be
\phi_{i} = \phi_{_{E_{i}}}(0) = {\sqrt{2m} x_2(E_i)\over 3g}\left[
{\cal E}\left( {x_1(E_i)^2\over x_2(E_i)^2}\right) - \sqrt{1-4E_ig^2}
{\cal K}\!\left({x_1(E_i)^2\over x_2(E_i)^2}\right) \right] - {\pi\over 4}\,.
\label{phii}
 \ee
The norm squared of (\ref{A7}) can be written in the form
  \begin{eqnarray}
|J^{12}(0)|^2 &=& {1\over 4} |A^*_{_{E_{1}}}A_{_{E_{2}}}|^2
\left\{ (c_{12}+c_{12}^{-1})^2 \mbox{cos}^{2}\phi_{1} +
(d_{12}+d_{12}^{-1})^2 \mbox{sin}^{2}\phi_{1} \right. \nonumber\\
&+& \left. \left[ (a_{12}+a_{12}^{-1}) \mbox{cos}\phi_{1}
\mbox{sin}\phi_{2} - (b_{12}+b_{12}^{-1}) \mbox{sin}\phi_{1}
\mbox{cos}\phi_{2} \right]^2 \right. \nonumber \\
&-& \left. \left[ (c_{12}+c_{12}^{-1}) \mbox{cos}\phi_1 \mbox{sin}\phi_2
- (d_{12}+d_{12}^{-1}) \mbox{sin}\phi_1 \mbox{cos}\phi_2
\right]^2 \right\} .
\label{A9}
  \end{eqnarray}
For fixed $E_1$ one may consider the right hand side of (\ref{A9}) as a
function of $E_2$. Using (\ref{abcd}),(\ref{phii}) and (\ref{delta}) it can be
shown that this function has a {\em global} minimum at $E_2=E_1$ where
trivially
$J^{11}(0)=J^{11}(\infty)$. Thus, for any other $E_2\neq E_1$ (in this energy
range) we have $|J^{12}(0)|>|J^{12}(\infty)|$ and (\ref{3.6}) does not hold.
Therefore ${\cal D}$ contains a single eigenstate in this $E$ range as well.

So far we saw that ${\cal D}$ contains at most a single eigenstate in each of
the energy ranges $E<V_{min}~,~V_{min}<E<V_{max}$ and $E>V_{max}$. Considering
now self adjointness of the Schr\"odinger operator in this finite dimensional
domain ${\cal D}$ we easily see that ${\cal D}$ is actually only {\em one
dimensional} (because otherwise $|J^{12}(0)| > |J^{12}(\infty)|$ for any pair
of
different such states) and the spectrum of $H$ shrinks to a single point
$E=E_0$.

For generalized tunneling wave functions which are not purely right (or left)
moving waves, the explicit form of the (non-zero) $J^{12}(0)$ is more
complicated and we currently do not have a complete proof that ${\cal D}$ is
one dimensional, but we present below indications that it is finite dimensional
at most. The WKB wave function to the right of the largest turning point
is
 \begin{eqnarray}
\Psi^{^{(WKB)}}_{E}(x>x_{_{>}}) &=& {1\over (E-V(x))^{1/4}} \left[
A_{_{E}} \mbox{exp} \left( i \int_{x_{_{>}}(E)}^x
\sqrt{2m(E-V(y))}dy \right) \right. \nonumber \\
&+& \left. B_{_{E}} \mbox{exp} \left(-i \int_{x_{_{>}}(E)}^x
\sqrt{2m(E-V(y))}dy \right) \right] ,
\label{A11}
  \end{eqnarray}
where $A_{_{E}}$ and $B_{_{E}}$ are complex amplitudes. For self adjoint
Hamiltonians with potentials satisfying (1.9) that are defined over an infinite
dimensional domain ${\cal D}$ the
spectrum is necessarily unbounded from below and from above. In this case
one can choose two eigenstates in ${\cal D}$ with energies $E_1$ and $E_2$
above $V_{max}$, such that $E_{2}/E_{1} \rightarrow \infty$. To leading order
in $(E_1/E_2)^{1\over 2} <<1 $ the current at $x=0$ is
  \be
J^{12}(0) \sim {1\over 2} \left( {E_{2}\over E_{1}} \right)^{1/4}
\left( A^*_{_{E_{1}}} A_{_{E_{2}}} - B^*_{_{E_{1}}} B_{_{E_{2}}}
+ B^*_{_{E_{1}}} A_{_{E_{2}}} - A^*_{_{E_{1}}} B_{_{E_{2}}} \right)\,.
\label{A12}
  \ee

On the other hand, the current at infinity is
  \be
J^{12}(x \rightarrow \infty) \rightarrow
A^*_{_{E_{1}}}A_{_{E_{2}}} e^{-i\delta_{12}}
- B^*_{_{E_{1}}}B_{_{E_{2}}} e^{i\delta_{12}} .
\label{A13}
  \ee
A generalized tunneling wave function is a wave function for which
$|A_{_{E}}| > |B_{_{E}}|$. For such wave functions the norm of
(\ref{A12}) diverges as $E_{2}/E_{1} \rightarrow \infty$, while
the norm of (\ref{A13}) is finite. Thus, $J^{12}(0)\neq J^{12}(\infty)$
and therefore, contrary to our initial assumption, the spectrum has to be
bounded from above.  Similarly, using (\ref{A4}) for negative energies we
find that $J^{12}(0)$ diverges when $|E_{2}|/|E_{1}| \rightarrow \infty$, and
the spectrum must be bounded from below as well. There must be a finite
minimum gap between energy levels in the band $0< E < {1\over 4g^2}$ and thus
the number of eigenstates is finite at most. Because this argument involves
the limit $E_2\rightarrow\infty$ it is insensitive to the details of $V(x)$
near $x=0$ as long as $V(x)$ and $V'(x)$ are regular at the origin.

$J^{12}(x=0)$ can be equal to $J^{12}(x \rightarrow \infty)$
for $E_{2}/E_{1} >> 1$ {\em only} if they are both zero.
Only in this case $J^{12}(x=0)$ will not diverge as $E_{2}/E_{1}
\rightarrow \infty$, and we can have a non-trivial
({\em infinite} dimensional)
domain for the self-adjoint Hamiltonian. This was studied in section 3.

\end{document}